\documentclass[12pt]{iopart}

\usepackage{epsfig}
\usepackage{graphicx}
\usepackage{dcolumn}
\usepackage{bm}

\newlength{\textwidthm}

\begin{document}

\title{Transmission through a defect in polyacene:
the extreme limit of ultra narrow graphene}

\author{N.~M.~R. Peres$^1$ and F. Sols$^2$}

\address{$^1$ Center of Physics  and  Department of
Physics, University of Minho, P-4710-057, Braga, Portugal}

\address{$^2$ Departamento de F\'{\i}sica de Materiales,
Universidad Complutense de Madrid, E-28040 Madrid, Spain}


\begin{abstract}
We compute the transmission of an electron through an impurity in 
polyacene. For simplicity the disorder is confined to a single unit 
cell. When the impurity preserves the inversion symmetry around the 
central axis, the scattering problem can be reduced to that of two 
independent chains with an alternating sequence of two types of 
atoms. An analytical expression for the transmission coefficient is 
derived. On-site and off-diagonal defects are considered and shown to 
display very different electron scattering properties.
\end{abstract}

\submitto{\JPCM}

\section{Introduction}
\label{Introd}

The study of polyacene started long ago in a pioneering paper by
Kivelson and Chapman.\cite{Kivelson83} The study performed by these
two authors was motivated by the investigations of Su, Schreiffer,
and Heeger on polyacetylene.\cite{Su79,Su80} They\cite{Kivelson83}
showed  that the band structure associated to the $\pi$-orbitals of
polyacene is made of four bands, with a valence and a conduction
bands touching at the limit of the one-dimensional Brillouin zone.
Since  polyacene can be considered the extreme case of a narrow
graphene\cite{novo1,pnas} ribbon it is instructive to compare the
band structure of graphene and polyacene. In both cases one has an
electron per carbon atom and in both cases the valence and
conduction band touch at the corner of the Brillouin zone. In
graphene, however, the bands are linear at the $\bm K$ and $\bm K'$
points of the Brillouin zone, whereas in polyacene the dispersion
relation around the momentum $k=\pi/a$ ($a$ the length of the unit
cell) is quadratic. As a consequence, graphene has a vanishing
density of states at the Fermi energy whereas polyacene displays a
square root singularity. It was therefore
speculated\cite{Kivelson83} that the ground state of polyacene could
support long range order such as ferromagnetic and superconducting
phases. It was further showed that phonon modes in polyacene can
lead to a Peierls distortion opening an energy gap at $k=\pi/a$.
\cite{Kivelson83,Bozovic85}

Using a Green's function method, Rosa and Melo\cite{Rosa88} studied
the density of states of several distorted configurations of
polyacene, including the case where a local defect was present. They
found that the optical response of polyacene should be very
different from that of polyacetylene. The inclusion of many-body
effects on the description of the ground state of polyacene was done
by Garcia-Bach {\it et al.}\cite{Garcia92} using a valence-bond
treatment. They found that the distorted ground state is degenerate.
Using a projector quantum Monte Carlo method, Srinivasan and
Ramasesha \cite{Srinivasan98} found, within the Hubbard model, that
electron-electron correlations tend to enhance the Peierls
instability. With the development of the density matrix
renormalization group, the exact study of the ground state of
quasi-one-dimensional electronic systems became available. Raghu
{\it et al.}\cite{Raghu02} studied the ground state of polyacene
using the Pariser-Parr-Pople Hamiltonian. As in previous studies
they found that strong electron-electron interactions can enhance
the dimerization. Using a configuration interaction technique, Sony
and Shukla \cite{Sony2007} studied the optical absorption and the
excited states of polyacenes also in the framework of the
Pariser-Parr-Pople Hamiltonian.

As interesting as the ground state nature of carbon polymers are
their transport properties. Of particular interest to us is the
effect of disorder on the electronic tunneling through a portion of
a disordered chain. Guinea and Verg\'es, \cite{Guinea87} using a
Green's function method, studied the local density of states and the
localization length of a one-dimensional chain coupled to small
pieces of laterally linked polymer. They showed that at the band
center there is an exact vanishing of the transmission coefficient
due to a local antiresonance. Sautet and Joachim \cite{Sautet88}
studied, within a one-dimensional tight-binding model, the effect of
a single impurity which would change both the on-site energy and the
hopping to the next-neighbor atoms. Mizes and Conwell \cite{Mizes91}
consider the effect of a single impurity in the square polymer,
showing that a change on the  on-site energy has a more pronounced
effect in reducing the transmission in the one-dimensional chain
than in the square polymer. At the end of the paper\cite{Mizes91}
these authors speculate that for polyacene there should be four
active scattering modes instead of two as in the square polymer. As
we show in this paper, this is not the case because the band
structure of polyacene is markedly different from that of the square
polymer. Gu {\it et al.} \cite{Gu92} generalized the study of Ref.
\cite{Mizes91} by including different types of obstacles as
scattering centers. Yu {\it el al.} \cite{Yu97} studied the
electronic transmission through a conjugate-oligomer using both mode
matching and Green's function methods. They found that the
transmission through the  conjugate-oligomer possesses several
transmission resonances. Farchioni {\it et al.}\cite{Farchioni}
studied the transport properties of emeraldine slats using a Green's
function method, by looking at the effect on the electronic
transmission of impurity dimers. The band structure of very narrow
carbon nanoribbons was studied by Ezawa. \cite{Ezawa06} The effect
on the conductance of the parity of the number of carbon rows across
a zig-zag ribbon has been studied by Akhmerov {\it et
al.}.\cite{Akhmerov07}

In this paper we study the effect of disorder on the electronic
transmission in polyacene. This system can be considered the most
extreme limit of narrow carbon nanoribbons,\cite{Ezawa06} and it
corresponds to an odd parity situation studied in Ref.
\cite{Akhmerov07}. We shall assume that the disorder is
limited to a single unit cell. Although this assumption can be
relaxed, it allows us however to obtain a full analytical expression
for the transmission coefficient. Following Ref.
\cite{Sautet88} we shall consider both on-site and hopping
disorder. The system is therefore characterized by two semi-infinite
perfect leads made of polyacene and a scattering region. The
Schr\"odinger equation has to be solved for the three regions and a
mode matching technique, first used by Ando \cite{Ando91} to study
the conductance of a square lattice in a magnetic field, will be
used here. Although, as discussed above, all the theoretical studies
point out that polyacene has a gap at the Fermi energy due to
phonons and interactions, here we will consider non-distorted
polyacene in the independent particle approximation. It should be
relatively simple to generalize the calculations below to include
the effect of a gap and the effect of electron-electron interactions
(at least at the mean field level), but the only expected
modification would be a small region near zero energy where the
conductance would be zero due to the presence of the gap. In
addition, the disorder could add states into  the gap of polyacene.

\section{Model for disordered polyacene}
\label{Model}

\begin{figure}[htf]
\begin{center}
\includegraphics*[angle=0,width=7cm]{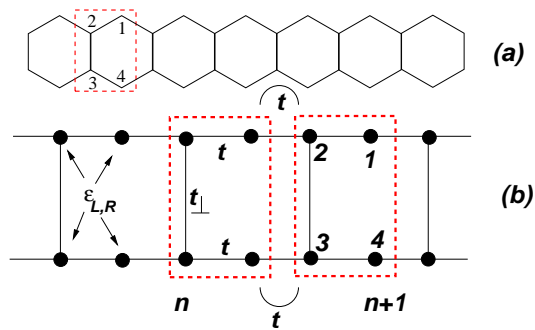}
\end{center}
\caption{(color online) Representation of polyacene in (a) and its
topologically equivalent lattice in (b). Each unit cell has four
atoms. The on-site energy is $\epsilon_{L,R}$. The hoping along the
chain is $t$ and in the perpendicular direction is $t_\perp$.}
\label{Fig_polyacene}
\end{figure}

\begin{figure}[htf]
\begin{center}
\includegraphics*[angle=0,width=7cm]{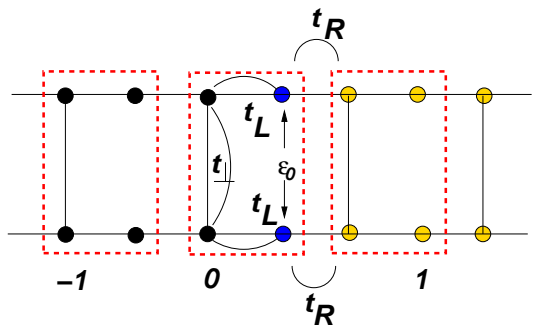}
\end{center}
\caption{(color online) Characterization of the $n=0$ unit cell,
where the disorder is located. Note that the on-site energies for
$n<0$ ($\varepsilon_L$) are different from those for $n>0$
($\varepsilon_R$). Also the on-site energy of atoms 1 and 4 in
the zero unit cell is different from the rest and
equals $\epsilon_0$.} \label{Fig_polyacene_zero_cell}
\end{figure}

The model for disordered polyacene can be written in general
terms as
\begin{eqnarray}
H&=&\sum_{n=-\infty}^\infty\sum_{i,j=1}^4
[\delta_{ij}\epsilon(n,i)\vert n,i\rangle\langle n,j\vert
\nonumber\\
&+&t_{\rm intra}(n,i,j) \vert n,i\rangle\langle n,j\vert
\nonumber\\
&+&t_{\rm inter}(n,i,j) \vert n,i\rangle\langle n+1,j\vert
]\,,
\end{eqnarray}
where $\epsilon(n,i)$ represents the on-site energy, $t_{\rm
intra}(n,i,j)$ the hopping between the atoms within the unit cell
$n$, and $t_{\rm inter}(n,i,j)$ the hopping between atoms in
neighboring unit cells. The sum over $n$ runs over the unit cells
and the sums over $i$ and $j$ run over the atoms in the unit cell.
In our model we shall assume that the on-site energies
$\epsilon(0,1)=\epsilon_0$ and $\epsilon(0,4)=\epsilon_0$ are
different from  those of the rest of the atoms in the lattice. The
on-site energies for atoms to the left of these two are all equal to
$\epsilon_L$ and, for those to the right, to $\epsilon_R$. This
difference could be due to a potential bias applied to the system.
The hopping parameters are  defined in Figs. \ref{Fig_polyacene} and
\ref{Fig_polyacene_zero_cell}. Only the hopping energies starting at
the atoms 1 and 4 of the zero unit cell are modified. They are named
$t_L$ and $t_R$. We further assume that the hopping energies between
atoms 2 and 3 within a given unit cell ($t_\perp$) can be different
from that between atoms 1 and 2, or 3 and 4  ($t$). The
eigenfunctions of this problem have the form
\begin{equation}
\vert\Psi\rangle = \sum_{n}\sum_jc(n,j)\vert n,j\rangle.
\end{equation}

As it is clear from Fig \ref{Fig_polyacene_zero_cell} the problem is
divided into three regions: (i) that to the left of the zero unit
cell, with $n=-\infty,\ldots,-1$, representing a semi-infinite
ordered lead; (ii) the disordered unit cell at $n=0$, which will
induce the scattering of the electrons; (iii) the right ordered
lead, for $n=1,\ldots,\infty$. We have therefore to solve the
problem in these three regions. The method we will use was developed
by Ando\cite{Ando91} in the context of quantum point contacts in a
magnetic field and generalized by Khomyakov {\it et al.}
\cite{{Khomyakov05}} to an arbitrary three-dimensional structure.

The strategy of solution is therefore the following, firstly the
problem is solved for the pure leads obtaining the eigenmodes. Then
the scattering problem is solved considering an incoming wave
described by one of the eigenmodes at a time. This allows for the
determination of the transmission matrix elements $t_{\mu\nu}$, and
from them the calculation of the conductance follows as
\cite{Buttiker85,Data}
\begin{equation}
G=\frac {2e^2} h \sum_{\mu\nu}\vert t_{\mu\nu}\vert^2\,,
\end{equation}
where the summation is over propagating channels only.

\section{Solution of the scattering problem}
\label{Scatt}

As explained above, the problem is separated into the solution in
the perfect leads and the solution in the scattering region. In what
follows we shall explain how to obtain those solutions.

\subsection{Scattering channels }

We study here the scattering channels, i.e. stationary solutions in
the asympotic leads assuming these are perfect and infinite. Let us
start with the left lead. The problem for the right one is solved
along the same lines but with $\epsilon_L$ replaced by $\epsilon_R$.
The tight-binding Hamiltonian for $m=-\infty,\ldots,-2$ has the form
\begin{equation}
-\bm B\bm c_{m-1}+(\bm I E-\bm H_L)\bm c_m - \bm B^\dag\bm
c_{m+1}=0\,, \label{TBlead}
\end{equation}
where $\bm c_m$ is a 4-vector containing the wave-function
coefficients of the unit cell $m$. The matrices $\bm B$ and $\bm
H_L$ are given by
\begin{equation}
\bm B=
\left(
\begin{array}{cccc}
 0 & 0 & 0 & 0\\
 t & 0 & 0 & 0\\
 0 & 0 & 0 & t\\
 0 & 0 & 0 & 0
\end{array}
\right)\,,
\label{B}
\end{equation}
and
\begin{equation}
\bm H_L=
\left(
\begin{array}{cccc}
 \epsilon_L & t & 0 & 0\\
 t &  \epsilon_L & t_\perp & 0\\
 0 & t_\perp &  \epsilon_L & t\\
 0 & 0 & t & \epsilon_L
\end{array}
\right)\,.
\label{H}
\end{equation}
It is important to note that the $\bm B$ matrix is singular, not
having an inverse. This will be important in what follows. As
explained by Ando, \cite{Ando91} the problem for the perfect lead
may be solved assuming a Bloch relation between the vectors $\bm
c_m$
\begin{equation}
\bm c_{m+1} = \lambda \bm c_m\,.
\label{bloch}
\end{equation}
For a general problem the vector $\bm c_m$ has dimension $M$ and the
solution of (\ref{TBlead}) can be obtained by transforming it
 into an ordinary
eigenvalue equation of dimension $2M$.\cite{Ando91} We look for
scattering states formed by an incoming wave approaching the
scattering region from the left and its resulting outgoing waves.
Whether these waves are propagating or evanescent states depends on
the nature of the eigenvalues. Propagating states always have
$\vert\lambda\vert =1$, whereas evanescent ones have $\vert \lambda
\vert\ne 1$. Another possibility is to transform Eq. (\ref{TBlead})
into a quadratic eigenvalue equation by repeatedly use of Eq.
(\ref{bloch}), resulting in
\begin{equation}
-\bm B\bm c_{m}+(\bm I E-\bm H_L)\lambda \bm c_m
- \bm B^\dag\lambda^2\bm c_{m}=0\,.
\label{TBlead2}
\end{equation}
As before, the determination of the eigenvalues of Eq.
(\ref{TBlead2}) will lead, for a problem of dimension $M$, to a
polynomial of order $2M$ whose roots are the sought eigenvalues.

Although we can attack the solution of the problem (\ref{TBlead2})
using the matrices given by Eqs. (\ref{B}) and (\ref{H}), it is
however convenient to perform a unitary transformation of Eq.
(\ref{TBlead2}). We define the unitary matrix  $\bm U$
\begin{equation}
\bm U= \frac{1}{\sqrt 2} \left(
\begin{array}{cccc}
 1 & 0 & 0 & 1\\
 0 &  1 & 1 & 0\\
 1 & 0 &  0 & -1\\
 0 & 1 & -1 & 0
\end{array}
\right)\,,
\label{U}
\end{equation}
and introduce the transformation $\bm {\tilde c}_m=\bm U\bm c_m$,
and $\bm {\tilde M} = \bm U \bm M\bm U^{-1}$. Upon this transformation,
the eigenproblem
(\ref{TBlead2}) is reduced to two block-diagonal quadratic
eigenvalue problems, since the resulting transforms of $\bm B$ and
$\bm H_L$ are factorized into two $2 \times 2$ matrices which we
denote generally as $\bm b$ and $\bm h_L$. Correspondingly, $\bm
{\tilde c}_{m}$ becomes the direct sum of two 2-vectors which we
generically refer to as $\bm u_m$ and which we assume normalized.
The resulting eigenvalue problem reads
\begin{equation}
-\bm b \bm  u_m+\lambda (\bm IE-\bm  h_L) \bm  u_m -\lambda^2\bm
b^\dag \bm  u_m=0\,, \label{litle}
\end{equation}
with
\begin{equation}
\bm b= \left(
\begin{array}{cc}
0 & 0 \\
t & 0
\end{array}
\right)
\end{equation}
and
\begin{equation}
\bm h_L= \left(
\begin{array}{cc}
\epsilon_L & t \\
t & \epsilon_L \pm t_\perp
\end{array}
\right)\,.\label{h-L}
\end{equation}
The decoupled two-dimensional problems describe propagation through
the even and odd modes with respect to the central axis of the
polyacene. The eigenvalue problem (\ref{litle}) has the same form as
that of a linear chain, with hopping energy $t$ between neighboring
atoms and with on-site energies alternating between $\epsilon_L$ and
$\epsilon_L\pm t_\perp$. As a conclusion, the scattering problem in
polyacene can be mapped into that of two independent one-dimensional
chains, contrary to the expectations of Mizes and Conwell.
\cite{Mizes91}

From the general discussion above one would expect that the
eigenvalue problem (\ref{litle}) would lead to a quartic polynomial
in $\lambda$. In fact because the matrix $\bm B$ has no inverse (as
happens for $\bm b$) the polynomial is only cubic in $\lambda$. One
of the solutions is  the  trivial one $\lambda=0$. This solution has
to be disregarded since it would produce a null wave function
everywhere. The other two solutions are
\begin{eqnarray}
\lambda &=&[(E-\epsilon_L)^2+t_\perp(E-\epsilon_L)-2t^2]/2t^2
\nonumber\\
&\pm& \frac 1 2
\sqrt{[(E-\epsilon_L)^2+t_\perp(E-\epsilon_L)-2t^2]^2/t^4-4}\,.
\label{lambda}
\end{eqnarray}
When the square root becomes imaginary, Eq. (\ref{lambda}) gives the
momentum of a propagating Bloch wave. If we now consider the case
$t_\perp\rightarrow -t_\perp$ [see Eq. (\ref{h-L})] two other
solutions are obtained.

The fact that we only have two solutions for each sign of $t_\perp$,
and not four, means that there are always two of the four expected
modes that do not contribute to the transport, not even as
evanescent waves. This result could have been anticipated if we had
considered the energy bands of perfect polyacene. \cite{Kivelson83}
These are given by
\begin{equation}
E-\epsilon_L=\pm \frac {t_\perp}{2} \pm \frac 1 2
\sqrt{t_\perp^2+8t^2[1+\cos(ka)]}\,, \label{bands}
\end{equation}
where $a$ is the length of the unit cell and $k\in [-\pi/a,\pi/a]$.
If we now try to solve for $k$ in Eq. (\ref{bands}) one finds that
for a given energy $E$ only two bands give a solution, being it real
or complex.

The velocity of the electrons in the modes is given by \cite{Khomyakov05}
\begin{equation}
v = -\frac{2a}{\hbar}\Im [\lambda \bm u^\dag \bm b^\dag \bm u]\,.
\end{equation}
For the present problem the velocity has a simple form given by
\begin{equation}
v=-\frac{2at}{\hbar}\Im [\lambda u_A u_B]\,, \label{veloc}
\end{equation}
where $u_A$ and $u_B$ are the components of $\bm u_m$, orbital $A$
resulting from the (symmetric or antisymmetric) linear combination
of sites 2 and 3 within the cell and orbital $B$ stemming from the
similar combination of sites 1 and 4 (see Fig. \ref{Fig_polyacene}).
These amplitudes are given by

\begin{equation}
u_A = \frac {t\vert 1+\lambda\vert}{\sqrt {(E-\epsilon_L)^2+
t^2\vert 1+\lambda\vert^2}}\,, \label{u1}
\end{equation}
and
\begin{equation}
u_B = \frac {(1+\lambda^\ast)(E-\epsilon_L)}{\vert 1 +\lambda\vert
\sqrt {(E-\epsilon_L)^2+ t^2\vert 1+\lambda\vert^2}}\,. \label{u2}
\end{equation}
Using Eqs. (\ref{u1}) and (\ref{u2}) the velocity (\ref{veloc}) reads
\begin{equation}
v=-\frac{2at}{\hbar} \frac {t(E-\epsilon_L)} {(E-\epsilon_L)^2+
t^2\vert 1+\lambda\vert^2}\Im\lambda\,. \label{vf}
\end{equation}
Equation (\ref{vf}) allows to identify the right and left moving modes
for a given energy $E$.
The right lead is solved in the same way with $\epsilon_L$ replaced by
$\epsilon_R$.

\subsection{The scattering region}

We now want to describe the scattering region. The Schr\"odinger equation
for the unit cell $m=-1$ has the same form as before, except that it couples to
$\bm c_0$. For the unit cell $m=0$ the Schr\"odinger equation is written
as

\begin{equation}
-\bm B\bm c_{-1}+(\bm I E-\bm H_{00}) \bm c_0
- \bm B^\dag_R\bm c_{1}=0\,
\label{cell0}
\end{equation}
with the matrices $\bm H_{00}$ and $\bm B_R$ given by
\begin{equation}
\bm H_{00}=
\left(
\begin{array}{cccc}
 \epsilon_0 & t_L & 0 & 0\\
 t_L &  \epsilon_L & t_\perp & 0\\
 0 & t_\perp &  \epsilon_L & t_L\\
 0 & 0 & t_L & \epsilon_0
\end{array}
\right)\,,
\label{H00}
\end{equation}
and
\begin{equation}
\bm B_R=
\left(
\begin{array}{cccc}
 0 & 0 & 0 & 0\\
 t_R & 0 & 0 & 0\\
 0 & 0 & 0 & t_R\\
 0 & 0 & 0  & 0
\end{array}
\right)\,.
\label{BR}
\end{equation}
For the unit cell $m=1$ the Schr\"odinger equation has the same form
as Eq. (\ref{TBlead}) except that $\bm B$ is replaced by $\bm B_R$,
$\epsilon_L$ is replaced by $\epsilon_R$ and it couples to $\bm
c_0$. As before we can perform a unitary transformation of the
Schr\"odinger equation leading to an effective $2\times 2$
Hamiltonian. The Hamiltonian for unit cell $m=-1$ has the same form
as Eq. (\ref{litle}). For the unit cell $m=0$ we obtain
\begin{equation}
-\bm  b\bm u_{-1}+ (\bm IE-\bm h_{00}) {\bm  u_0}- \bm b_R^\dag\bm
u_{1}=0\,,
\end{equation}
and for $m=1$
\begin{equation}
-\bm  b_R\bm u_0+ (\bm IE-\bm h_R) {\bm  u_1}- \bm b^\dag\bm
u_{2}=0\,.
\end{equation}
The matrix $\bm  b_R$ has the same form as $\bm  b$ with $t$
replaced by $t_R$. The matrix $\bm h_R$ is obtained from $\bm h_L$
replacing $\epsilon_L$ by $\epsilon_R$. The matrix $\bm h_{00}$ is
given by
\begin{equation}
\bm h_{00}= \left(
\begin{array}{cc}
\epsilon_0 & t_L \\
t_L & \epsilon_L \pm t_\perp
\end{array}
\right)\,.
\end{equation}
Since the full problem factorizes into two block-diagonal problems
and, furthermore, because in the leads, for each $2\times 2$ block,
only one mode is active, the scattering takes place without mode
mixing. This is a considerable simplification over the general
approach of Refs. \cite{Ando91,Khomyakov05}.

If we define $\bm u^{\pm}_m$ as the amplitude at cell $m$ of the
Bloch wave propagating to the right ($+$) or left ($-$) in perfect
polyacene, the following relation holds:
\begin{equation}
\bm  u^{\pm}_{-m+1}=\lambda^{\pm}\bm  u^{\pm}_{-m} \label{LCmm1} \,
,
\end{equation}
where $\lambda^{\pm}=1/\lambda^{\mp}$ is given in Eq. (\ref{lambda})
for the asympotic left lead. Using Eq. (\ref{LCmm1}) we can write
\begin{eqnarray}
\bm  u_{-2}= \left( \lambda^- - \lambda^+\right)\bm u^+_{-1}+
\lambda^+ \bm u_{-1} \,. \label{c2m}
\end{eqnarray}
\begin{figure}[htf]
\begin{center}
\includegraphics*[angle=0,width=6cm]{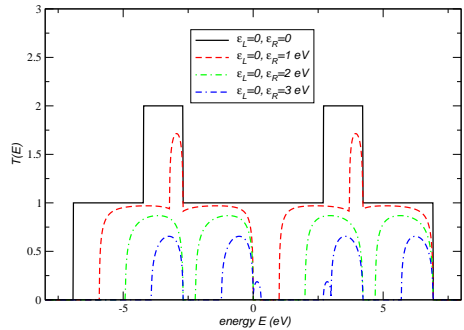}
\end{center}
\caption{(color on line) Representation of the transmission $T(E)$
for different values of $\epsilon_R$. The parameters are
$t=t_\perp=t_L=t_R=-2.7$ {\ttfamily eV}, $\epsilon_0=\epsilon_L=0$.
The solid line corresponds to perfect polyacene.}
\label{Fig_polyacene_step}
\end{figure}
\begin{figure}[htf]
\begin{center}
\includegraphics*[angle=0,width=6cm]{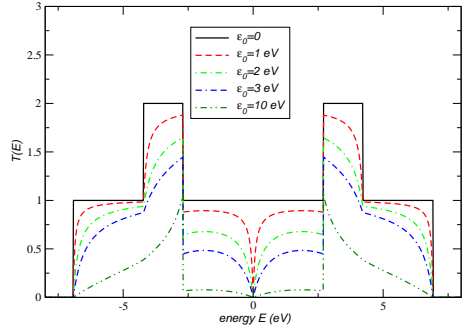}
\end{center}
\caption{(color on line)
Representation of the transmission $T(E)$ for different values of
$\epsilon_0$.
 The parameters are $t=t_\perp=t_L=t_R=-2.7$ {\ttfamily eV},
$\epsilon_L=\epsilon_R=0$. The solid line corresponds to perfect
polyacene.} \label{Fig_polyacene_local_energy}
\end{figure}
The boundary conditions require the specification of $\bm u^+_{-1}$,
which will represent an incoming wave function in one of the modes
of the left lead. On the right lead we have only a scattered wave
propagating to the right,
 we thus write
\begin{eqnarray}
\bm  u_2 = \lambda^+\bm u_1\,. \label{c2p}
\end{eqnarray}
Using Eqs. (\ref{c2m}) and (\ref{c2p}), the determination of the
wave function on the unit cells $m=-1,0,1$ reduces to the resolution
of the following system of linear equations
\begin{equation}
[\bm IE-\bm  h_L-\lambda^+\bm  b]\bm  u_{-1}- \bm b^\dag \bm u_0 = (
\lambda^- - \lambda^+) \bm u^+_{-1}\,, \label{LS1}
\end{equation}
\begin{equation}
-\bm  b\bm u_{-1}+ (\bm IE-\bm h_{00}) {\bm  u_0}- \bm b^\dag_R\bm
u_{1}=0\,, \label{LS2}
\end{equation}
\begin{equation}
-\bm  b_R\bm  u_0 +(\bm IE-\bm  h_R-\bm b^\dag\lambda^+) \bm
u_1=0\,. \label{LS3}
\end{equation}
Once the system is solved the vector $\bm  u_1$ is determined. We
can then write $\bm  u_1$ as
\begin{equation}
\bm  u_1 = \bm u^+_1 \tau_{\alpha} \,,
\end{equation}
where $\alpha$ is an explicit index to label the (odd or even) mode.
The physical transmission matrix element is given by\cite{Data}
\begin{equation}
t_{\alpha}=\sqrt{\frac {v_R}{v_L}}\tau_{\alpha}\,,
\end{equation}
where $v_{L/R}$ represents the velocity of the considered mode in
the left/right lead, given by Eq. (\ref{vf}). The total conductance
is obtained by summing the contributions from the symmetric and the
antisymmetric mode,
\begin{equation}
G=\frac {2e^2}{h}T(E) =\frac {2e^2}{h}\sum_{\alpha=1}^{2}\vert
t_{\alpha}\vert^2\,,
\end{equation}
assuming that both modes are propagating waves. Solving explicitly
the linear system of equations defined above, the expression for
$\tau_{\alpha}$ is obtained,
\begin{equation}
\tau_{\alpha}=qt^3  t_Lt_R u^+_A\lambda^-\frac \eta D\,,
\label{taui}
\end{equation}
where
 \begin{equation}
q=\lambda^- - \lambda^+ \,,
 \end{equation}
\begin{equation}
\eta =  \frac {E-\epsilon_R}{u_B^+}\, ,
\end{equation}
and $u^+_A,u^+_B$ are the $m$-independent parts of the $A$ and $B$
components of $\bm  u_m^+$ in the left and right leads,
respectively.
 The denominator $D$ reads
\begin{eqnarray}
D&=&
-t^2_L[t^2+(t^2-E(E+t_\perp)+\epsilon_L(2E+t_\perp-\epsilon_L))\lambda^-_L]
\nonumber\\
&\times&
[t^2-E(E+t_\perp)+\epsilon_R(2E+t_\perp-\epsilon_R)+t^2\lambda^+_R]
\nonumber\\
&+& (E+t_\perp-\epsilon_L)[-t^2+(-2t^2+E(E+t_\perp)-\epsilon_L
(2E+t_\perp-\epsilon_L))\lambda^-_L]
\nonumber\\
&\times &
[t^2_R(-E+\epsilon_R)+(E-\epsilon_0)(-t^2+E(E+t_\perp) -
\epsilon_R(2E+t_\perp-\epsilon_R)-t^2\lambda^+_R)]\,.
\end{eqnarray}

We are now in position to study the transmission coefficient
$T(E)$. In Figure \ref{Fig_polyacene_step} we study the transmission
of the electrons in a situation that mimics that of a potential
step in ordinary quantum mechanics problems. The potential barrier
is created by having the sites at the right of the unit cell
$n=0$ at a different energy from those at the left. For definiteness
we consider $\epsilon_L=0$. When $\epsilon_L=\epsilon_R$, $T(E)$
has a step structure, represented by the solid line
in Fig. \ref{Fig_polyacene_step},
 due to the existence of two possible conducting
channels. These two conducting channels are those associated with
the two effective one-dimensional chains. For energies where the two
channels of the two chains are propagating states, one obtains
$T(E)=2$; when only one channel is active one obtains $T(E)=1$.
Having now $\epsilon_L\ne\epsilon_R$ induces some back scattering of
the electrons at the interface, reducing the value of $T(E)$. When
the difference between $\epsilon_L$  and $\epsilon_R$ increases  we
see the appearance of zones of zero transmission, while this is
always nonzero for $\epsilon_L=\epsilon_R$.

\begin{figure}[htf]
\begin{center}
\includegraphics*[angle=0,width=6cm]{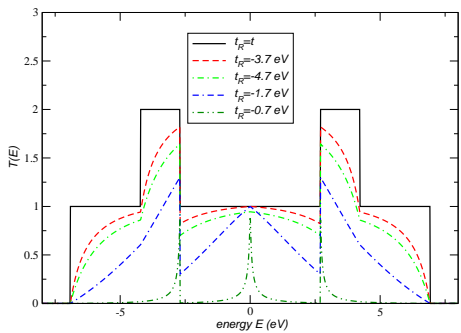}
\end{center}
\caption{(color on line)
Representation of the transmission $T(E)$ for different values of
the hopping $t_R$ and $t_L$.
 The parameters are $t=t_\perp=-2.7$ {\ttfamily eV},
$\epsilon_L=\epsilon_R=\epsilon_0=0$, $t_R=t_L$. The solid line
corresponds to perfect polyacene.} \label{Fig_polyacene_hopping}
\end{figure}

In Figure \ref{Fig_polyacene_local_energy} we study the effect of
changing the value of $\epsilon_0$ relatively to that of
$\epsilon_L$ and $\epsilon_R$. Overall there is an effect of
diminishing $T(E)$. If the difference between $\epsilon_0$ and
$\epsilon_L$ and $\epsilon_R$ is not very large the $T(E)$ curve
follows closely that for the non-disordered case. There is however
an exception at zero energy, where  $T(E)$ tends to zero, leading to
a totally reflecting barrier. This can be due to a local
antiresonance as in Refs. \cite{Guinea87,sols89}.

On the contrary, if the disorder is induced by changing the hopping,
the behavior of $T(E)$ shows almost perfect transmission at $E=0$,
as can be seen in Fig. \ref{Fig_polyacene_hopping}. This effect is
particularly clear when $t_L$ and $t_R$, become very small. From Eq.
(\ref{taui}) we see that $T(E)$ will vanish as the fourth power of
$t_L$ ($t_L=t_R$); nevertheless close to $E=0$ the system has a
strong enhancement of its transmission. Away from this special point
the curves for $T(E)$ are similar to those of Fig.
\ref{Fig_polyacene_local_energy} for on-site disorder.

In the cases of Figs.  \ref{Fig_polyacene_local_energy} and
\ref{Fig_polyacene_hopping} the conductance of the system would
exhibit conductance oscillations as long as the scattering strength
of the defect is not too strong.

\section{Conclusions}

In this paper we have studied analytically the electronic 
transmission of polyacene due to on-site and off-diagonal disorder. 
We have shown that the system is equivalent to two decoupled linear 
chains of atoms, with alternating on-site energies and a constant hopping
parameter. Scattering occurs in the system without mode mixing if 
inversion symmetry around the central axis is preserved. We find that 
on-site and hopping disorder have markedly distinct scattering 
properties close to zero energy. Whereas for on-site disorder the 
transmission decreases strongly close to zero energy, for hopping 
disorder the transmission is enhanced. If the system opens up a gap 
at the Fermi energy, we expect these same characteristics to occur 
near the gap edge.

\section*{Acknowledgments}

NMRP acknowledges the financial support from POCI 2010 (Portugal)
through Grant PTDC/FIS/64404/2006. FS has been supported by MEC
(Spain) under Grants FIS2004-05120 and FIS2007-65723.

\end{document}